\documentclass[article,twocolumn, floatfixnofootinbib]{revtex4}
\usepackage{filecontents}
\usepackage{hyperref}
\usepackage{graphicx} 
\usepackage{subfig}
\usepackage{color}

\newcommand{\ket}[1]{\left| #1 \right\rangle}
\newcommand{\bra}[1]{\left\langle #1 \right|}

\newcommand{\be}{\begin{equation}}
\newcommand{\ee}{\end{equation}}
\newcommand{\bea}{\begin{eqnarray}}
\newcommand{\eea}{\end{eqnarray}}

\usepackage{dcolumn}
\usepackage{bm}
\usepackage{amssymb,amsmath}
\usepackage{color}
\usepackage{float}
\usepackage{tikz}
\usetikzlibrary{arrows}

\definecolor{DarkGreen}{rgb}{0,0.6,0.2}

\begin{document}
\title{Strong coupling optical spectra in dipole-dipole interacting optomechanical Tavis-Cummings models}
\author{$^{1,2}$Imran M. Mirza}
\affiliation{ $^{1}$Department of Physics, University of Michigan, Ann Arbor, MI 48109-1040, USA\\
$^{2}$Oregon Center for Optics and Department of Physics\\University of Oregon, 
Eugene, OR 97403}

\begin{abstract}
We theoretically investigate the emission spectrum of an \emph{optomechanical Tavis-Cummings model}: two dipole-dipole interacting atoms coupled to an optomechanical cavity (OMC). In particular, we study the influence of \emph{dipole-dipole interaction (DDI)} on the single-photon spectrum emitted by this hybrid system in the presence of a strong atom-cavity as well as strong optomechanical interaction (hereinafter called the \emph{strong-strong coupling}). We also show that our analysis is amenable to inclusion of mechanical losses (under the \emph{weak mechanical damping limit}) and single-photon loss through spontaneous emission from the two-level emitters under a \emph{non-local Lindblad model}.
\end{abstract}
\maketitle
The hybrid quantum systems have gathered a considerable attention recently due to their (proposed) ability to outperform disparate tasks in quantum information processing and quantum computation which can not be accomplished by the individual components utilized in such systems \cite{kurizki2015quantum,schliesser2011hybrid}. The hybrid atom-optomechanics \cite{rogers2014hybrid, wallquist2010single} is a captivating example in this regard. On one hand, quantum optomechanics \cite{aspelmeyer2014cavity} has opened up new venues in both applications and foundations of quantum theory \cite{kleckner2008creating, stannigel2012optomechanical,gavartin2012hybrid}. Whereas, on the other hand the field of cavity quantum electrodynamics (CQED) \cite{haroche1989cavity} has its own history and a wide range of applications in controlled light matter interactions \cite{mckeever2003experimental,hartmann2006strongly} (to name a very few).\\
A unique combination of optomechanics and CQED, qubit assisted optomechanical hybrid systems are particularly fascinating as they provide a platform to investigate a wealth of phenomena which can arise as a result of atom-light and light-mechanics coherent interactions at the quantum level. These effects are particularly useful in building emergent quantum-enhanced technologies. For some recent and compelling efforts in this area we direct readers to the references \cite{de2011entanglement,yin2013large}.
From the fundamental as well as application point of view, it is crucial to understand the spectral properties of hybrid atom-optomechanical systems under various parameter regimes. In this context, the stationary \cite{nunnenkamp2011single} as well as time-dependent spectrum \cite{liao2012spectrum,imran2014single} of an empty OMC and the real-time spectrum of an OMC containing one atom \cite{mirza2015real} has already been investigated. Here we move forward to discuss a novel situation (which to our knowledge has not been studied yet), in which two dipole-dipole interacting (DDI) atoms are trapped inside an optomechanical cavity and we investigate the spectral characteristics of the system valid under a strong-strong coupling regime. In comparison to our previous work on single atom-OMC systems we now consider a full non-local Lindblad model, as it is known from the work of Walls, Cresser and Joshi et.al that the phenomenonlogical addition of individual dissipative terms in the master equation (local Lindblad model) can lead to erroneous results for quantum systems mutually coupled with arbitrary coupling strengths \cite{walls1970higher,cresser1992thermal,joshi2014markovian}.\\
 We find that the DDI can bring novel features to the optical spectra of hybrid atoms-OMC system. For example, the DDI can modify and gives us more control over the Rabi peak positions as well as their relative heights in the spectra. In this perspective, we notice that the DDI acts like a positive atom-cavity detuning and has an indirect influence on the mechanical side-bands as well. Finally, we also introduce finite mechanical damping in our analysis and spontaneous emission events from two-level emitters.
\begin{figure}[t]
\includegraphics[width=3.4in,height=1.6in]{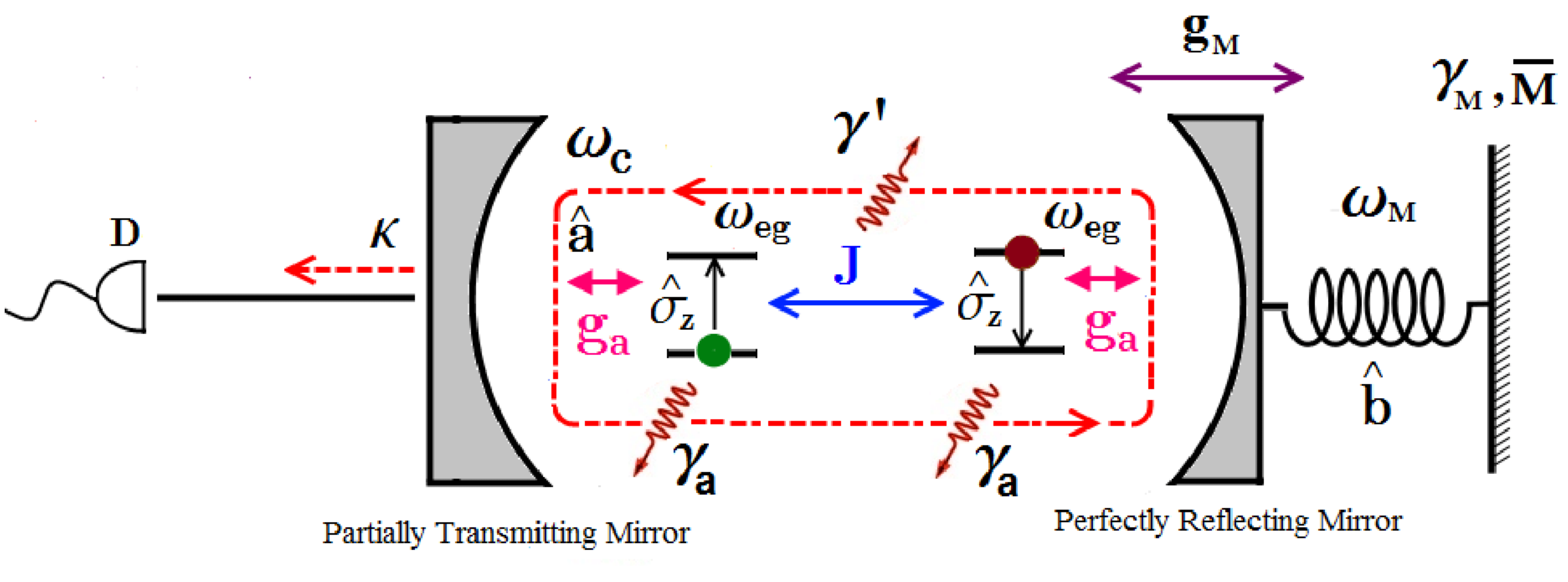}
\captionsetup{
  format=plain,
  margin=1em,
  justification=raggedright,
  singlelinecheck=false
}
 \caption{Two dipole-dipole interacting atoms coupled to an optomechanical cavity. Initially, one of the atoms is in an excited state and mechanical oscillator is assumed to be in a zero phonon state. The optical spectrum recorded through the detector $D$ is used to probe the signatures of atom-atom, atom-cavity and optomechanical coherent interactions.}\label{Fig1}
\end{figure}
As shown in Fig.~1, system under consideration consists of two fixed DDI two-level atoms trapped in a Fabry-Per\'ot optomechnical cavity. For simplicity both atoms are assumed to be identical with a ground ($\ket{g}$) and an excited state ($\ket{e}$) and transition frequency and spontaneous emission rate given by $\omega_{eg}$ and $\gamma_{a}$ respectively. Exchange of photon from one atom to the other gives rise to a position dependent interaction which is equivalent to the dipole-dipole interaction between two classical dipoles \cite{nicolosi2004dissipation} (hence the name DDI). The strength of DDI $J$ can be expressed as: $J=3/4(\gamma_{0}c^{3}_{0}/\omega^{3}_{eg}r^{3})$, where $\gamma_{0}$ is the free space spontaneous emission rate, $c_{0}$ is the speed of light in free space and $r$ is the inter-atomic separation.
The mechanical part of the OMC is a perfectly reflecting ultra-thin mirror which is modeled as a quantum harmonic oscillator with equilibrium frequency $\omega_{M}$ and phonon destruction represented by the operator $\hat{b}$. Optical cavity is assumed to have a single isolated resonant mode with frequency $\omega_{c}$ while the annihilation of photons in the cavity is described by operator $\hat{a}$. The atom-field interaction strength is characterized through $g_{a}$ parameter and the rate $g_{M}$ indicates the optomechanical interaction strength. The Hamiltonian of the system under rotating wave approximation (RWA) and in a frame rotating with frequency $\omega_{c}$ can be expressed as:
\begin{equation}
\begin{split}
&\hat{H}_{sys}= -\hbar\sum_{i=1,2}\Bigg[\Delta_{ac}\hat{\sigma}^{\dagger}_{i}\hat{\sigma}_{i}-g_{a}(\hat{a}^{\dagger}\hat{\sigma}_{i}+\hat{a}\hat{\sigma}^{+}_{i})\Bigg]\\
&+\hbar J(\hat{\sigma}^{+}_{1}\hat{\sigma}_{2}+\hat{\sigma}^{+}_{2}\hat{\sigma}_{1})
+\hbar\omega_{M}\hat{b}^{\dagger}\hat{b}-\hbar g_{M}\hat{a}^{\dagger}\hat{a}(\hat{b}^{\dagger}+\hat{b}),
\end{split}
\end{equation}
Here $\Delta_{ac}=\omega_{eg}-\omega_{c}$. In applying DDI to CQED part, we have kept in view the frequencies involved in typical CQED experiments where $\omega_{eg},\omega_{c}>>\kappa$ ($\omega_{eg}$ and $\omega_{c}\sim$ few tens of GHz while the cavity leakage rate $\kappa$ and spontaneous emission rate $\gamma_{a}$ are usually around 10MHz \cite{birnbaum2005photon,guerlin2007progressive}). Non-vanishing commutation relations are given by: $[\hat{\sigma}_{+},\hat{\sigma}_{-}]=\hat{\sigma}_{z}$, $[\hat{a},\hat{a}^{\dagger}]=1$ and $[\hat{b},\hat{b}^{\dagger}]=1$. \\
To study the open system dynamics of the setup shown in Fig.~1, we employ the following Lindblad Markovian master equation:
\begin{equation}
\begin{split}
&\frac{d\hat{\rho}_{s}(t)}{dt} = \hat{\mathcal{L}}_{sys}[\hat{\rho}_{s}]+\frac{\gamma_{a}}{2}\sum_{i=1,2}\hat{\mathcal{L}}_{\hat{\sigma}_{i}}[\hat{\rho}_{s}]+\frac{\gamma^{'}_{a}}{2}(\hat{\mathcal{L}}_{\hat{\sigma}_{1},\hat{\sigma}_{2}}[\hat{\rho}_{s}]\\
&+\hat{\mathcal{L}}_{\hat{\sigma}_{2},\hat{\sigma}_{1}}[\hat{\rho}_{s}])+\frac{\kappa}{2}\hat{\mathcal{L}}_{\hat{a}}[\hat{\rho}_{s}]+\frac{(2\beta)^{2}\gamma_{M}}{ln(1+1/\overline{M})}\hat{\mathcal{L}}_{\hat{a}^{\dagger}\hat{a}}[\hat{\rho}_{s}]\\
&+\frac{\gamma_{M}}{2}(\overline{M}+1)\hat{\mathcal{L}}_{\hat{b}-\beta\hat{a}^{\dagger}\hat{a}}[\hat{\rho}_{s}]
+\frac{\gamma_{M}\overline{M}}{2}\hat{\mathcal{L}}_{\hat{b}^{\dagger}-\beta\hat{a}^{\dagger}\hat{a}}[\hat{\rho}_{s}],
\end{split}
\end{equation}
here $\hat{\mathcal{L}}_{sys}[\hat{\rho}_{s}]\equiv\frac{-i}{\hbar}[\hat{H}_{sys},\hat{\rho}_{s}(t)]$,$\hat{\mathcal{L}}_{\hat{\sigma}_{1},\hat{\sigma}_{2}}[\hat{\rho}_{s}]\equiv(2\hat{\sigma}_{1}\hat{\rho}_{s}(t)\hat{\sigma}^{\dagger}_{2}-\hat{\sigma}^{\dagger}_{1}\hat{\sigma}_{2}\hat{\rho}_{s}-\hat{\rho}_{s}(t)\hat{\sigma}^{\dagger}_{1}\hat{\sigma}_{2})$ and $\hat{\mathcal{L}}_{\hat{O}}[\hat{\rho}_{s}]\equiv2\hat{O}\hat{\rho}_{s}(t)\hat{O}^{\dagger}-\hat{O}^{\dagger}\hat{O}\hat{\rho}_{s}(t)-\hat{\rho}_{s}(t)\hat{O}^{\dagger}\hat{O}$. $\beta\equiv g_{M}/\omega_{M}$ while $\gamma_{M}$ and $\overline{M}$ are the mechanical damping rate and average thermal phonon number, respectively. $\gamma^{'}_{a}$ is a position dependent cooperative decay rate, which originates from the fact that both atoms are coupled to a common vacuum bath \cite{alharbi2010deterministic}. \\
It is worthwhile to note that above master equation has a non-local Lindblad structure for dissipation from OMC due to a strong optomechanical interaction ($g_{M}\sim\omega_{M}$) as also pointed out in the references \cite{hu2015quantum,holz2015suppression}. In particular in Ref.\cite{holz2015suppression}, Holz et. al has applied the exact same master equation to study Rabi oscillations' suppression in strong hybrid atom optomechanics. Here atom-cavity losses are treated locally due to the cavity QED (or circuit QED) parameters used typically i.e. $g_{a}<<\omega_{eg},\omega_{c}$ but $g_{a}>\kappa$ and the rotating wave approximation for cavity QED part holds. For the spectrum calculations we take the infinite limit of the Eberly and Wodkiewicz (E\&W) \cite{eberly1977time} definition in which the optical spectrum is interpreted as filtered counting rate :
\begin{equation}
\begin{split}
&N(t;\Delta,\Gamma)=\\
&\kappa\Gamma^{2}\int\int e^{-(\Gamma-i\Delta)(t-t^{'})}e^{-(\Gamma+i\Delta)(t-t^{''})}\langle\hat{a}^{\dagger}(t^{'})\hat{a}(t^{''})\rangle dt^{'}dt^{''},
\end{split}
\end{equation}
where filter cavity is assumed to have a Lorentzian spectral profile with a fixed bandwidth $\Gamma$ but a variable frequency $\omega$ $(\Delta=\omega-\omega_{c})$. \\
In our previous studies \cite{imran2014single,mirza2015real} we have presented a dressed state analysis with a multiple phonon restriction. Such a situation is an artificial scenario and was adopted merely for the sake of simplicity. Here we present a full dressed state analysis which is valid for any arbitrary phonon number.\\
 To this end, we neglect all sources of decoherence and concentrate on the system Hamiltonian alone. As it is difficult to perform a dressed state analysis with two DDI atoms \cite{zhang2014effects}, we first transform the system Hamiltonian into an effective single-atom Hamiltonian through the transformation $\hat{U}=e^{-\frac{\pi}{4}(\hat{\sigma}^{\dagger}_{1}\hat{\sigma}_{2}-\hat{\sigma}^{\dagger}_{2}\hat{\sigma}_{1})}$, under a low atomic excitation assumption \cite{nicolosi2004dissipation}. The new Hamiltonian then represents two fictitious atoms: one atom (with shifted frequency $\Delta_{ac}-J$) is coupled to the cavity field (with an effective rate $\sqrt{2}g_{a}$) while the other atom is completely cavity decoupled. Next, we transform the optomechanical part of the state of the system by applying the so called ``polaron transformation": $\ket{\tilde{m}}=e^{-\frac{g_{M}}{\omega_{M}}(\hat{b}-\hat{b}^{\dagger})}\ket{m}$. As a result the transformed system Hamiltonian splits into two parts:
\begin{equation}
\begin{split}
&\hat{H}_{0}=-\hbar(\Delta_{ac}-J)\hat{\sigma}^{\dagger}\hat{\sigma}
+\hbar\omega_{M}\hat{b}^{\dagger}\hat{b}-\hbar g_{M}\hat{a}^{\dagger}\hat{a}(\hat{b}^{\dagger}+\hat{b})\\
&\hat{H}_{I}=-\sqrt{2}g_{a}(\hat{a}^{\dagger}\hat{\sigma}+\hat{a}\hat{\sigma}^{+}).
\end{split}
\end{equation}
The presence of $\hat{H}_{I}$ mixes the unperturbed eigenstates to form the following dressed states:
\begin{subequations}
\begin{eqnarray}
\ket{\Psi^{(+)}_{m}(t)}={\rm sin}\Theta\ket{g,1,\tilde{m}}+{\rm cos}\Theta\ket{e,0,m},\\
\ket{\Psi^{(-)}_{m}(t)}=-{\rm cos}\Theta\ket{g,1,\tilde{m}}+{\rm sin}\Theta\ket{e,0,m},
\end{eqnarray}
\end{subequations}
while ${\rm tan 2}\Theta=\frac{2\sqrt{2}g_{a}}{\tilde{\Delta}_{ac}+\frac{g^{2}_{M}}{\omega_{M}}}$ while $\tilde{\Delta}_{ac}\equiv\Delta_{ac}-J$ and corresponding eigenvalues are:
\begin{equation}
\begin{split}
&\varepsilon^{(\pm)}_{n,m}=\frac{\tilde{\Delta}_{ac}}{2}+m\omega_{M}-\frac{g^{2}_{M}}{2\omega_{M}}\pm\frac{1}{2}\sqrt{(\tilde{\Delta}_{ac}+\frac{g^{2}_{M}}{\omega_{M}})^{2}+8g^{2}_{a}}.
\end{split}
\end{equation}
Note that the eigenvalues varies with the optomechanical coupling ($g_{M}$) in a quadratic manner, which is the exact same dependence one obtains for the empty OMC case \cite{nunnenkamp2011single} also. In Fig.~2 we have plotted the optical spectrum to examine the influence of DDI. We note the following points:\\
\begin{figure}[h]
\includegraphics[width=2.6in, height=1.8in]{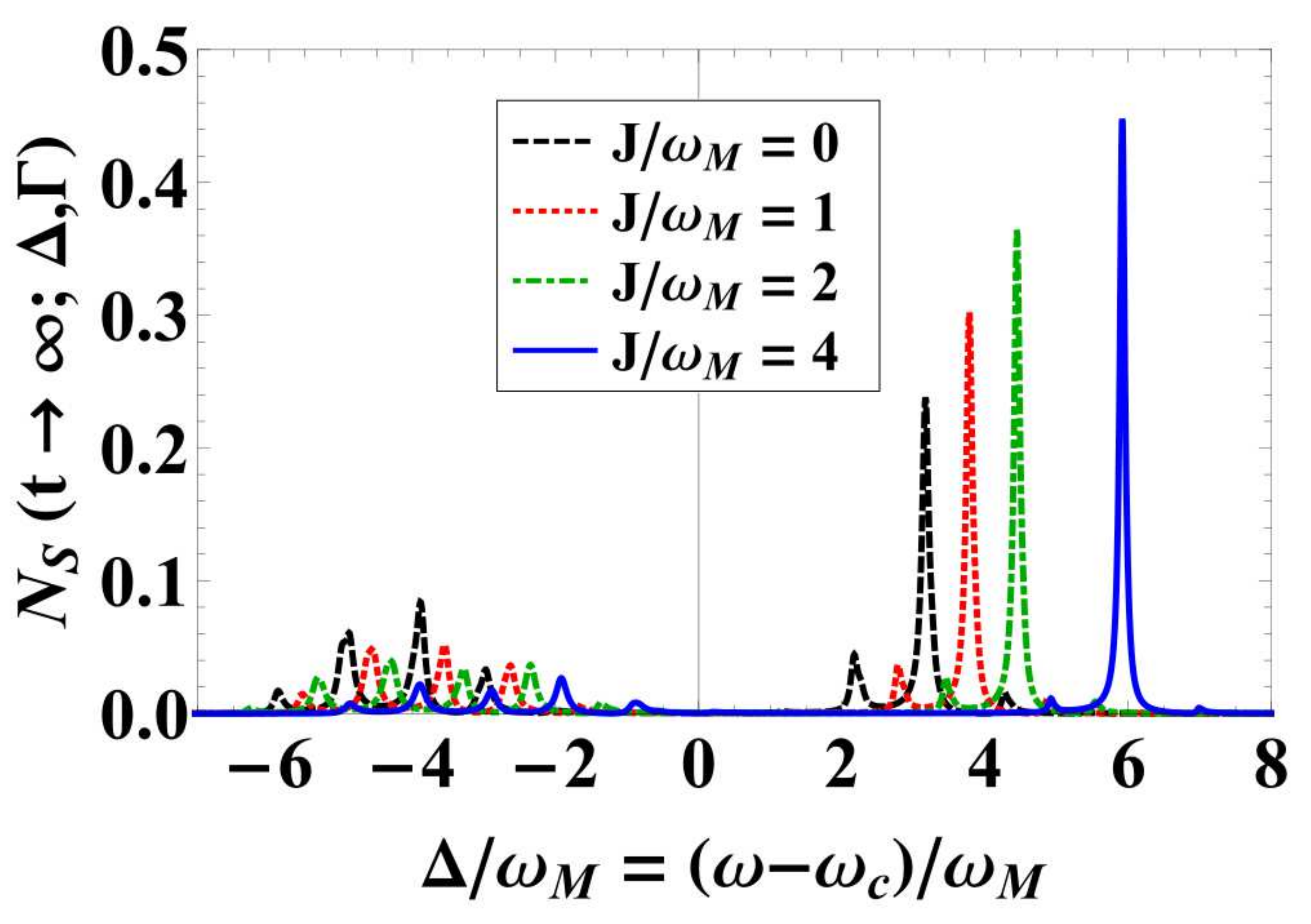}
\captionsetup{
  format=plain,
  margin=1em,
  justification=raggedright,
  singlelinecheck=false
}
\caption{Single-photon stationary spectrum emitted by DDI atoms-OMC system in the strong-strong coupling regime. Parameters used in this and next two figures are: $g_{a}/\omega_{M} = 2.4, g_{M}/\omega_{M} = 1.2, \kappa/\omega_{M} = 0.2$, $\Gamma/\omega_{M} = 0.01, \gamma_{a}/\omega_{M} = 0.05, \Delta_{ac}/\omega_{M} =0$, $\gamma_{M}/\omega_{M} =0$, and $\overline{M} = 0$.}
\end{figure}
$\bullet$ First of all when $J=0$, the separation between two major peaks is $\sim 6.8\omega_{M}$, which is consistent with the prediction made by dressed state picture, in which case $\varepsilon=\varepsilon^{(+)}_{n,m}-\varepsilon^{(-)}_{n,m}=\hbar\sqrt{(g^{2}_{M}/\omega_{M})^{2}+8g^{2}_{a}}$. It is worthwhile to mention that the optomechanical coupling modifies the standard vacuum Rabi splitting (of $8g^{2}_{a}$) by a factor of $g^{2}_{M}/\omega^{2}_{M}$ here.\\
$\bullet$ Asymmetry in the peak heights is linked with the transition from any one of the excited state to ground state $\ket{g,0,0}$. For example, transition from $\ket{\Psi^{(+)}_{m}}$ to $\ket{g,0,0}$ is given by: $|\bra{g,0,0}\sqrt{\kappa}\hat{a}\ket{\Psi^{(+)}_{m}}|^{2}=\kappa {\rm sin^{2}}\Theta(\frac{e^{-g^{2}_{M}/2\omega^{2}_{M}}}{m!})(\frac{-g_{M}}{\omega_{M}})^{2m}|\mathbb{L}^{m}_{0}(g^{2}_{M}/\omega^{2}_{M})|^{2}$, where $\mathbb{L}^{m}_{m^{'}}(x)$ are the associated Laguerre polynomials. Similarly, for transition from $\ket{\Psi^{(-)}_{m}}\rightarrow\ket{g,0,0}$ we obtain:
$|\bra{g,0,0}\sqrt{\kappa}\hat{a}\ket{\Psi^{(-)}_{m}}|^{2}
=\kappa {\rm cos^{2}}\Theta(\frac{e^{-g^{2}_{M}/2\omega^{2}_{M}}}{m!})(\frac{-g_{M}}{\omega_{M}})^{2m}$
Since these matrix elements are level dependent, so its not surprising that there is an asymmetry in peak heights.\\
$\bullet$ The presence of the mechanical side bands which are occurring at an integer multiple of $\omega_{M}$ can also be understood by noticing the dependence of $\varepsilon^{(\pm)}_{n,m}$ (Eq.~13) on $m\omega_{M}$ term. \\
$\bullet$ Next, the separation between the main peaks tend to grow as $J$ is increased.  This is because (and as seen from the expression $\varepsilon ^{(\pm)}_{n,m}$) that for $J\neq0$, the distance between the two peaks is a monotonically increasing function of $J$ (separation goes like $\sqrt{(J+g^{2}_{M}/\omega_{M})^{2}+8g^{2}_{a})}$).\\
$\bullet$ Finally, enhancing the $J$ value also causes further enhancement in a major peak asymmetry. This is due to the fact that with raising DDI the function $sin\Theta(cos\Theta)$ value decreases (increases).\\
The effect of varying atom-cavity detuning on the single-photon spectrum for the case of zero and non-zero dipole-dipole coupling are presented in Fig.~3 and 4, respectively. Three cases of positive, negative and zero detuning are considered.\\
\begin{figure}[h]
\includegraphics[width=2.6in, height=1.8in]{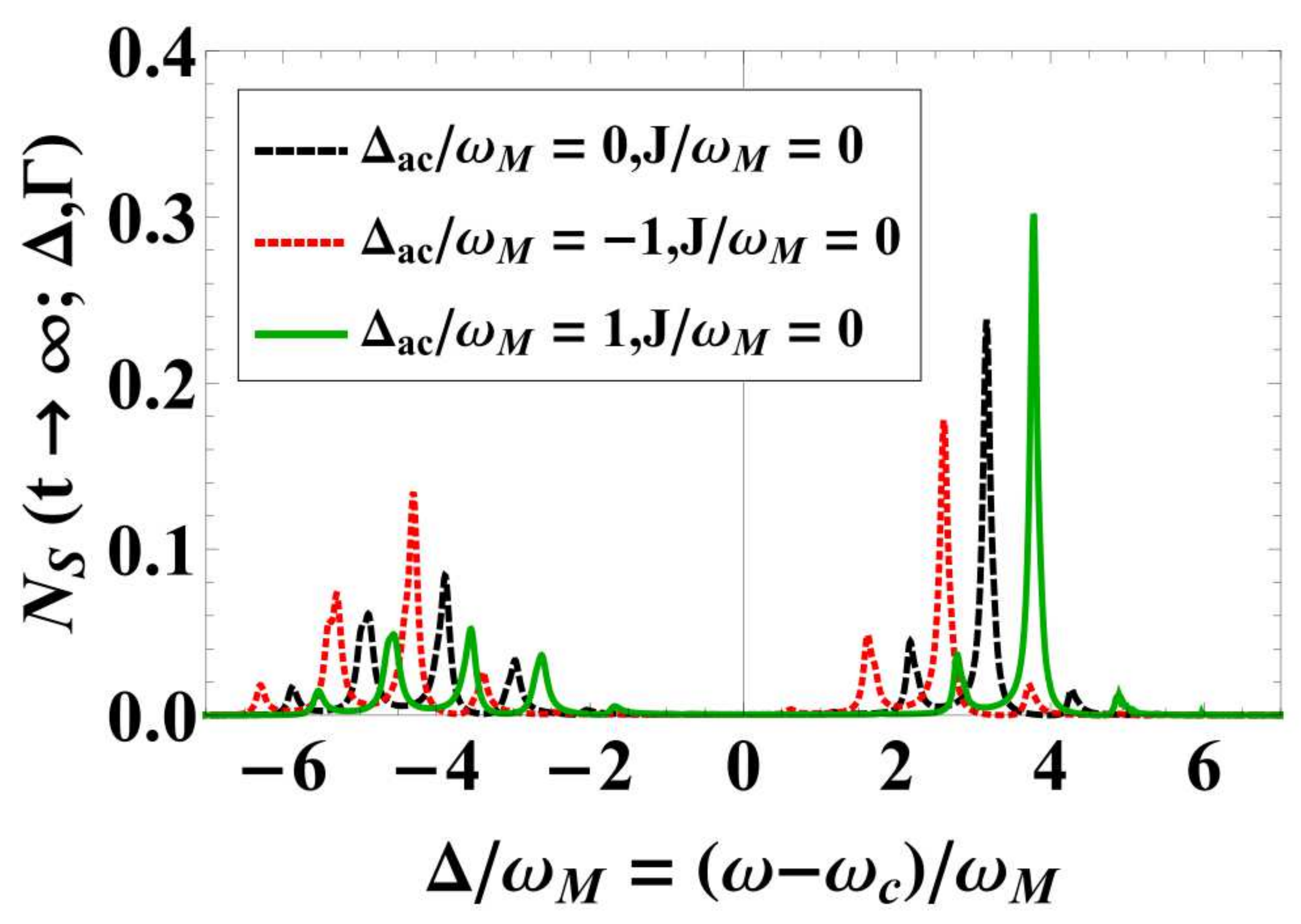}
\captionsetup{
  format=plain,
  margin=1em,
  justification=raggedright,
  singlelinecheck=false
}
\caption{Single-photon stationary spectrum plotted to exhibit the effect of varying the atom-cavity detuning on the emitted spectrum.}
\end{figure}  
\begin{figure}
\includegraphics[width=2.6in, height=1.8in]{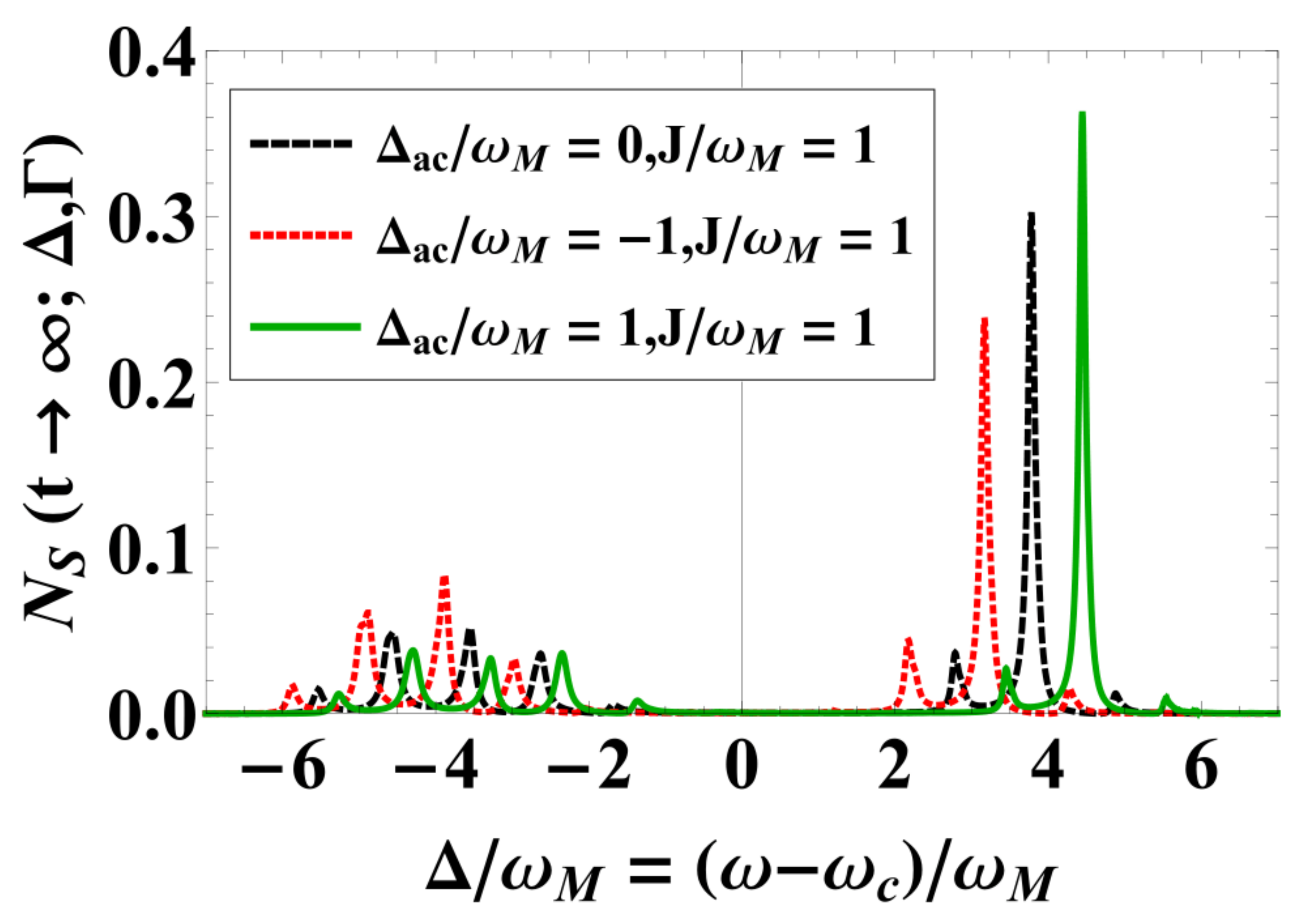}
\captionsetup{
  format=plain,
  margin=1em,
  justification=raggedright,
  singlelinecheck=false
}
\caption{ This spectrum is plotted to show how the effect of changing atom-cavity detuning and DDI between atoms both effect cooperatively and alter the spectrum.}
\end{figure}
$\bullet$ We notice that for positive and negative detunings spectrum shift around the on resonance peak position. The reason being, the dependence of peak positions on the term $\Delta_{ac}-J$ in the normal-mode energies' expression ($\varepsilon^{(\pm)}_{n,m}$ ). With increasing $\Delta_{ac}$ (from -1 to +1) one finds the major peak separation to enhance slightly as predicted by $\Delta\varepsilon^{(\pm)}=\sqrt{(\Delta_{ac}+g^{2}_{M}/\omega_{M})^{2}+8g^{2}_{a}}$.\\
$\bullet$ Interestingly, in Fig.~4 we observe that for increasing major peak separation, positive detuning plays the same role as played by the dipole-dipole interaction. When both $\Delta_{ac}$ and $J$ being positive (green thick curve in the figure), we have the largest peak separation and peak asymmetry, while when $\Delta_{ac}<0$ but $J>0$ (red dotted curve) both $\Delta$ and $J$ have a some sort of cancellation effect and peak separation and asymmetry becomes the smallest.
\begin{figure}
\begin{center}
\includegraphics[width=6cm,height=4.5cm]{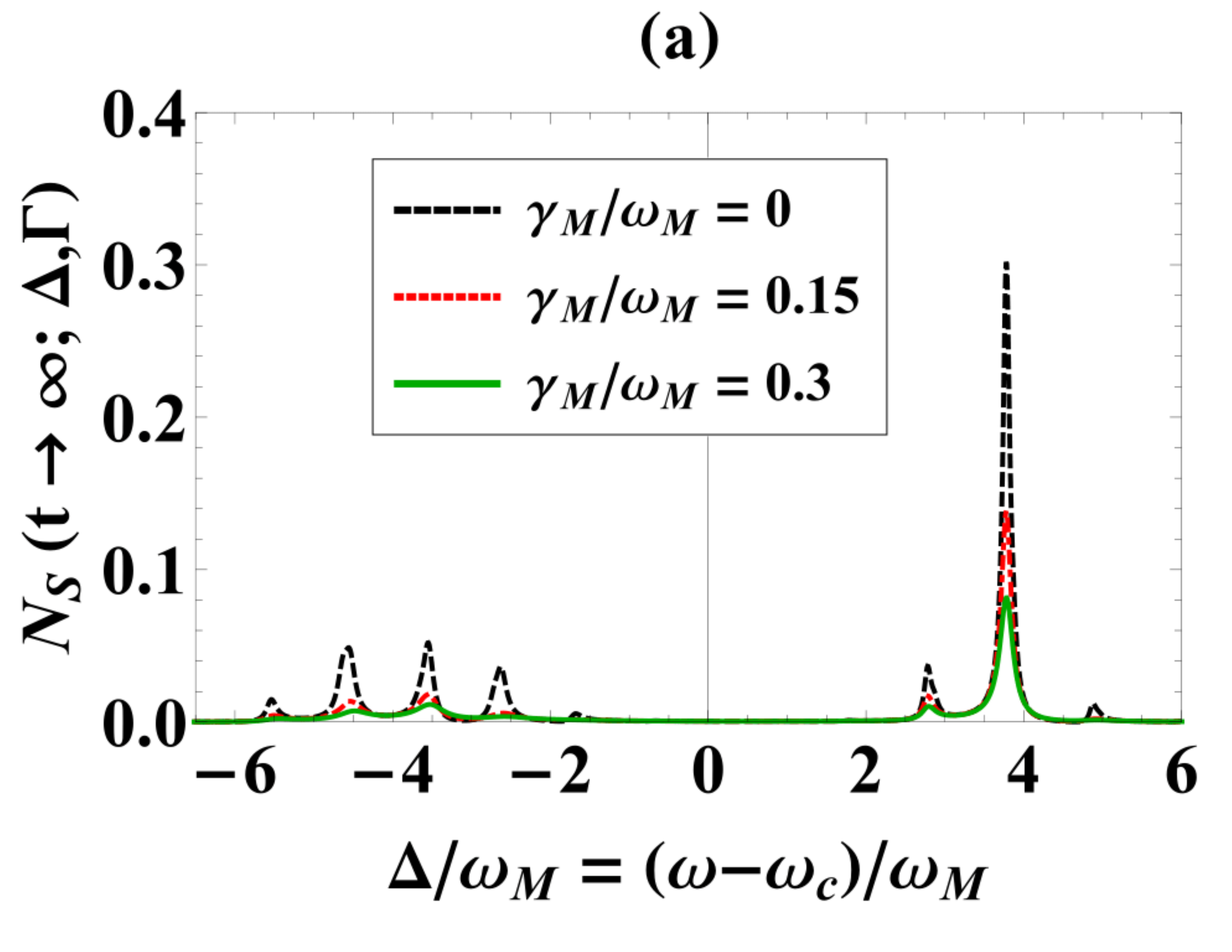}\\ 
\includegraphics[width=6cm,height=4.5cm]{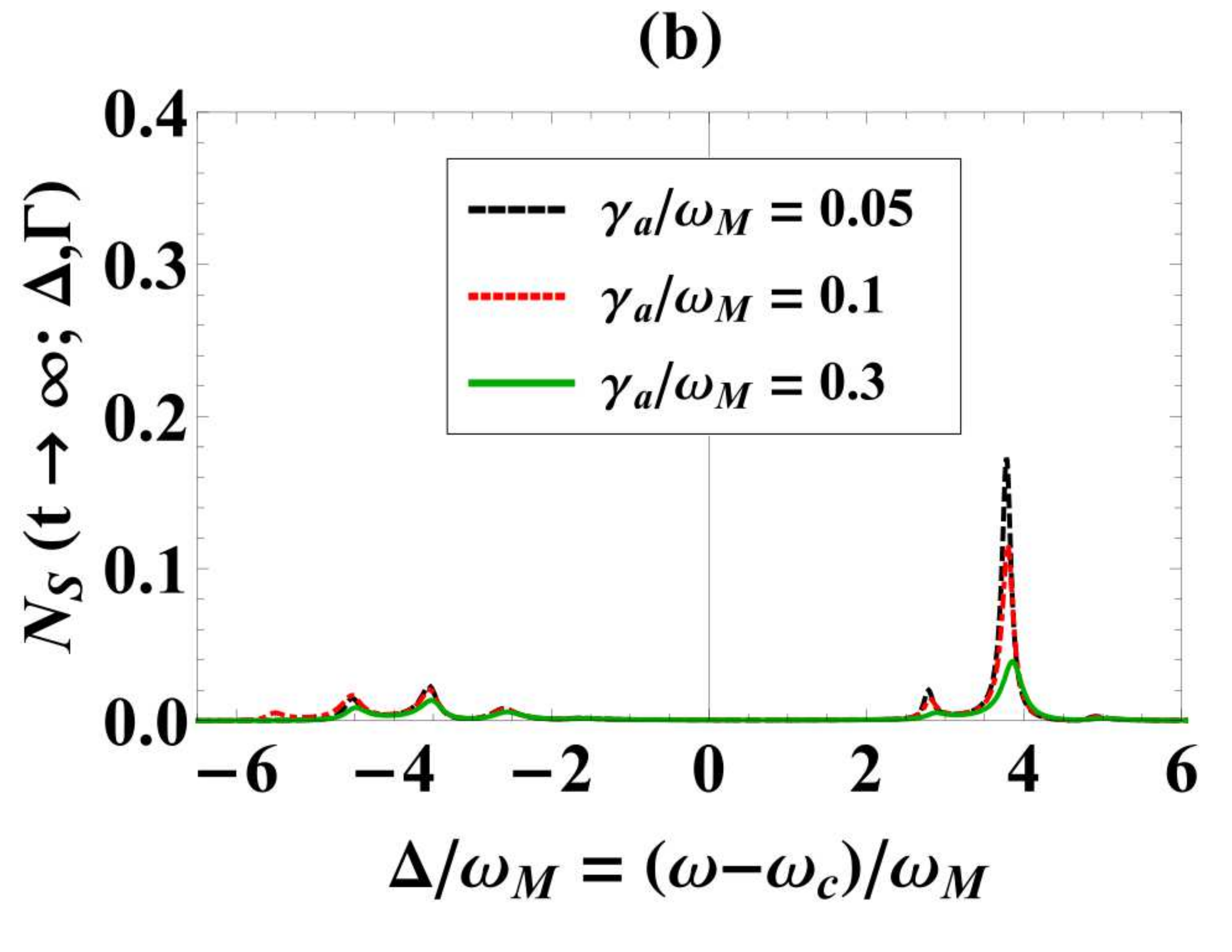}
\captionsetup{
  format=plain,
  margin=1em,
  justification=raggedright,
  singlelinecheck=false
}
\caption{Effect of varying (a) $\gamma_{M}$ and (b) $\gamma_{a}$ on the single photon spectrum. For plot (a) $\gamma_{a}/\omega_{M} = 0.1$ and for part (b) $\gamma_{M}/\omega_{M} = 0.05$. In both plots we have considered a strong-strong coupling (with remaining parameters same as used in Fig.~2) and an on resonance situation with $J=1\omega_{M}$.}\label{Fig5}
\end{center}
\end{figure}
Although in Fig.~2-4 we have already taken a very small value of the spontaneous emission rate $\gamma_{a}/\omega_{M}=0.05$, in this section we'll focus on varying $\gamma_{a}$ to higher values and also incorporate finite mechanical damping. To this end, we apply the full master equation described in Eq.~2 and consider movable mirror to be coupled with a memory-less (i.e. Markovian) mechanical heat bath. Initially, the mechanical bath is assumed to be in a thermal state with Boltzmann occupancy factor:
\begin{equation}\label{IBD}
p_{m_{0}}(t=0)=\frac{\overline{M}^{m_{0}}}{(1+\overline{M})^{m_{0}+1}},
\end{equation}
while label $m_{0}$ and $\overline{M}$ are respectively the initial and average thermal phonon numbers. In Fig.~5 (a) we have plotted the spectrum with varying $\gamma_{M}$ values ($\Delta_{ac}, J $ and $\overline{M}$ are fixed) under a weak damping limit ($\overline{M}<<1$ and $\kappa>>\gamma_{M}$). As compared to Fig.~2, we notice that the application of full master equation now (where optical and mechanical losses can influence each other) results in an enhancement of mechanical side bands heights. This also effects the major Rabi peaks and an overall shifting of all resonances appearing on $-\Delta$-axis towards right occur (due to $\hat{b}-\beta\hat{a}^{\dagger}\hat{a}$ terms in the master equation). We have not applied the dressed state analysis when mechanical losses are present, yet exact peak locations can still be obtained by setting real part of the poles in the optical spectrum equal to zero. For a full treatment on obtaining peak positions using this pole technique in single photon empty OMC systems see reference \cite{liao2012spectrum}.
We notice that as we increase the mechanical damping the heights of the peaks tend to decrease. For $\gamma_{M}=0.3\omega_{M}$ we also notice a slight broadening of peaks. This causes the smallest side-bands (lying outside $\Delta=-6\omega_{M}$ and $6.5\omega_{M}$ range) to vanish completely from the spectrum. Here we would like to mention a relevant study \cite{nunnenkamp2011single} on single-photon optomechanics (empty OMC) where a similar blurring of the spectrum into wide thermal background has also been reported. \\
Finally, in Fig.~5(b) we vary the atomic spontaneous emission rate ($\gamma_{a}$) to higher values (between $0.05\omega_{M}$ to $0.3\omega_{M}$ for fixed $\gamma_{M}$). Like Fig.~5(a) and in comparison to Fig.~2, we again note peak enhancement of mechanical side-bands and shifting of red side bands (and the corresponding Rabi peaks) due to application of Eq.~2. We also notice that similar to $\gamma_{M}$ variation, the effect of enhancing $\gamma_{a}$ also causes the peak height reduction. This happens due to the fact that with increasing $\gamma_{a}$ photon loss into the unwanted spontaneous emission channel also grows which causes a decrease in the photonic filtered counting rate (spectrum) and hence peak height reduction results.\\

In conclusion, we have studied the emission spectra of a hybrid DDI atom-OMC system in a strong-strong coupling regime under a non-local Lindblad model. We noticed with an increase in the dipole-dipole interaction, two major vacuum Rabi splitted peaks become more asymmetric as well as separated. The same effect is produced by the positive detuning and infact both positive detuning and the DDI act collectively to enhance these features. We found that under weak excitation limit and applying the Polaron transformation one can perform a dressed state analysis of the problem which excellently explains both peaks locations and asymmetry under the variation of various parameters involved. The inclusion of mechanical damping necessitates the use a master equation with non-local Lindblad structure for optical losses and mechanical damping. These losses result in a considerable peak height reduction as well as broadening of the resonances' bandwidth.\\

\bibliography{Article}

\begin{thebibliography}{28}
\expandafter\ifx\csname natexlab\endcsname\relax\def\natexlab#1{#1}\fi
\expandafter\ifx\csname bibnamefont\endcsname\relax
  \def\bibnamefont#1{#1}\fi
\expandafter\ifx\csname bibfnamefont\endcsname\relax
  \def\bibfnamefont#1{#1}\fi
\expandafter\ifx\csname citenamefont\endcsname\relax
  \def\citenamefont#1{#1}\fi
\expandafter\ifx\csname url\endcsname\relax
  \def\url#1{\texttt{#1}}\fi
\expandafter\ifx\csname urlprefix\endcsname\relax\def\urlprefix{URL }\fi
\providecommand{\bibinfo}[2]{#2}
\providecommand{\eprint}[2][]{\url{#2}}

\bibitem[{\citenamefont{Kurizki et~al.}(2015)\citenamefont{Kurizki, Bertet,
  Kubo, M{\o}lmer, Petrosyan, Rabl, and Schmiedmayer}}]{kurizki2015quantum}
\bibinfo{author}{\bibfnamefont{G.}~\bibnamefont{Kurizki}},
  \bibinfo{author}{\bibfnamefont{P.}~\bibnamefont{Bertet}},
  \bibinfo{author}{\bibfnamefont{Y.}~\bibnamefont{Kubo}},
  \bibinfo{author}{\bibfnamefont{K.}~\bibnamefont{M{\o}lmer}},
  \bibinfo{author}{\bibfnamefont{D.}~\bibnamefont{Petrosyan}},
  \bibinfo{author}{\bibfnamefont{P.}~\bibnamefont{Rabl}}, \bibnamefont{and}
  \bibinfo{author}{\bibfnamefont{J.}~\bibnamefont{Schmiedmayer}},
  \bibinfo{journal}{, Proc. Natl. Acad. Sci. USA}
  \textbf{\bibinfo{volume}{112}}, \bibinfo{pages}{3866} (\bibinfo{year}{2015}).

\bibitem[{\citenamefont{Schliesser and
  Kippenberg}(2011)}]{schliesser2011hybrid}
\bibinfo{author}{\bibfnamefont{A.}~\bibnamefont{Schliesser}} \bibnamefont{and}
  \bibinfo{author}{\bibfnamefont{T.~J.} \bibnamefont{Kippenberg}},
  \bibinfo{journal}{, Physics} \textbf{\bibinfo{volume}{4}},
  \bibinfo{pages}{97} (\bibinfo{year}{2011}).

\bibitem[{\citenamefont{Rogers et~al.}(2014)\citenamefont{Rogers, Lo~Gullo,
  De~Chiara, Palma, and Paternostro}}]{rogers2014hybrid}
\bibinfo{author}{\bibfnamefont{B.}~\bibnamefont{Rogers}},
  \bibinfo{author}{\bibfnamefont{N.}~\bibnamefont{Lo~Gullo}},
  \bibinfo{author}{\bibfnamefont{G.}~\bibnamefont{De~Chiara}},
  \bibinfo{author}{\bibfnamefont{G.~M.} \bibnamefont{Palma}}, \bibnamefont{and}
  \bibinfo{author}{\bibfnamefont{M.}~\bibnamefont{Paternostro}},
  \bibinfo{journal}{, Quant. Meas. and Quant. Metrol.}
  \textbf{\bibinfo{volume}{2}} (\bibinfo{year}{2014}).

\bibitem[{\citenamefont{Wallquist et~al.}(2010)\citenamefont{Wallquist,
  Hammerer, Zoller, Genes, Ludwig, Marquardt, Treutlein, Ye, and
  Kimble}}]{wallquist2010single}
\bibinfo{author}{\bibfnamefont{M.}~\bibnamefont{Wallquist}},
  \bibinfo{author}{\bibfnamefont{K.}~\bibnamefont{Hammerer}},
  \bibinfo{author}{\bibfnamefont{P.}~\bibnamefont{Zoller}},
  \bibinfo{author}{\bibfnamefont{C.}~\bibnamefont{Genes}},
  \bibinfo{author}{\bibfnamefont{M.}~\bibnamefont{Ludwig}},
  \bibinfo{author}{\bibfnamefont{F.}~\bibnamefont{Marquardt}},
  \bibinfo{author}{\bibfnamefont{P.}~\bibnamefont{Treutlein}},
  \bibinfo{author}{\bibfnamefont{J.}~\bibnamefont{Ye}}, \bibnamefont{and}
  \bibinfo{author}{\bibfnamefont{H.}~\bibnamefont{Kimble}}, \bibinfo{journal}{,
  Phys. Rev. A} \textbf{\bibinfo{volume}{81}}, \bibinfo{pages}{023816}
  (\bibinfo{year}{2010}).

\bibitem[{\citenamefont{Aspelmeyer et~al.}(2014)\citenamefont{Aspelmeyer,
  Kippenberg, and Marquardt}}]{aspelmeyer2014cavity}
\bibinfo{author}{\bibfnamefont{M.}~\bibnamefont{Aspelmeyer}},
  \bibinfo{author}{\bibfnamefont{T.~J.} \bibnamefont{Kippenberg}},
  \bibnamefont{and}
  \bibinfo{author}{\bibfnamefont{F.}~\bibnamefont{Marquardt}},
  \bibinfo{journal}{, Rev. of Mod. Phys.} \textbf{\bibinfo{volume}{86}},
  \bibinfo{pages}{1391} (\bibinfo{year}{2014}).

\bibitem[{\citenamefont{Kleckner et~al.}(2008)\citenamefont{Kleckner, Pikovski,
  Jeffrey, Ament, Eliel, Van Den~Brink, and
  Bouwmeester}}]{kleckner2008creating}
\bibinfo{author}{\bibfnamefont{D.}~\bibnamefont{Kleckner}},
  \bibinfo{author}{\bibfnamefont{I.}~\bibnamefont{Pikovski}},
  \bibinfo{author}{\bibfnamefont{E.}~\bibnamefont{Jeffrey}},
  \bibinfo{author}{\bibfnamefont{L.}~\bibnamefont{Ament}},
  \bibinfo{author}{\bibfnamefont{E.}~\bibnamefont{Eliel}},
  \bibinfo{author}{\bibfnamefont{J.}~\bibnamefont{Van Den~Brink}},
  \bibnamefont{and}
  \bibinfo{author}{\bibfnamefont{D.}~\bibnamefont{Bouwmeester}},
  \bibinfo{journal}{, New J. Phys.} \textbf{\bibinfo{volume}{10}},
  \bibinfo{pages}{095020} (\bibinfo{year}{2008}).

\bibitem[{\citenamefont{Stannigel et~al.}(2012)\citenamefont{Stannigel, Komar,
  Habraken, Bennett, Lukin, Zoller, and Rabl}}]{stannigel2012optomechanical}
\bibinfo{author}{\bibfnamefont{K.}~\bibnamefont{Stannigel}},
  \bibinfo{author}{\bibfnamefont{P.}~\bibnamefont{Komar}},
  \bibinfo{author}{\bibfnamefont{S.}~\bibnamefont{Habraken}},
  \bibinfo{author}{\bibfnamefont{S.}~\bibnamefont{Bennett}},
  \bibinfo{author}{\bibfnamefont{M.~D.} \bibnamefont{Lukin}},
  \bibinfo{author}{\bibfnamefont{P.}~\bibnamefont{Zoller}}, \bibnamefont{and}
  \bibinfo{author}{\bibfnamefont{P.}~\bibnamefont{Rabl}}, \bibinfo{journal}{,
  Phys. Rev. Lett.} \textbf{\bibinfo{volume}{109}}, \bibinfo{pages}{013603}
  (\bibinfo{year}{2012}).

\bibitem[{\citenamefont{Gavartin et~al.}(2012)\citenamefont{Gavartin, Verlot,
  and Kippenberg}}]{gavartin2012hybrid}
\bibinfo{author}{\bibfnamefont{E.}~\bibnamefont{Gavartin}},
  \bibinfo{author}{\bibfnamefont{P.}~\bibnamefont{Verlot}}, \bibnamefont{and}
  \bibinfo{author}{\bibfnamefont{T.}~\bibnamefont{Kippenberg}},
  \bibinfo{journal}{, Nature nanotech.} \textbf{\bibinfo{volume}{7}},
  \bibinfo{pages}{509} (\bibinfo{year}{2012}).

\bibitem[{\citenamefont{Haroche and Kleppner}(1989)}]{haroche1989cavity}
\bibinfo{author}{\bibfnamefont{S.}~\bibnamefont{Haroche}} \bibnamefont{and}
  \bibinfo{author}{\bibfnamefont{D.}~\bibnamefont{Kleppner}},
  \bibinfo{journal}{, Phys. Today} \textbf{\bibinfo{volume}{42}},
  \bibinfo{pages}{24} (\bibinfo{year}{1989}).

\bibitem[{\citenamefont{McKeever et~al.}(2003)\citenamefont{McKeever, Boca,
  Boozer, Buck, and Kimble}}]{mckeever2003experimental}
\bibinfo{author}{\bibfnamefont{J.}~\bibnamefont{McKeever}},
  \bibinfo{author}{\bibfnamefont{A.}~\bibnamefont{Boca}},
  \bibinfo{author}{\bibfnamefont{A.~D.} \bibnamefont{Boozer}},
  \bibinfo{author}{\bibfnamefont{J.~R.} \bibnamefont{Buck}}, \bibnamefont{and}
  \bibinfo{author}{\bibfnamefont{H.~J.} \bibnamefont{Kimble}},
  \bibinfo{journal}{, Nature} \textbf{\bibinfo{volume}{425}},
  \bibinfo{pages}{268} (\bibinfo{year}{2003}).

\bibitem[{\citenamefont{Hartmann et~al.}(2006)\citenamefont{Hartmann, Brandao,
  and Plenio}}]{hartmann2006strongly}
\bibinfo{author}{\bibfnamefont{M.~J.} \bibnamefont{Hartmann}},
  \bibinfo{author}{\bibfnamefont{F.~G.} \bibnamefont{Brandao}},
  \bibnamefont{and} \bibinfo{author}{\bibfnamefont{M.~B.}
  \bibnamefont{Plenio}}, \bibinfo{journal}{, Nature Physics}
  \textbf{\bibinfo{volume}{2}}, \bibinfo{pages}{849} (\bibinfo{year}{2006}).

\bibitem[{\citenamefont{De~Chiara et~al.}(2011)\citenamefont{De~Chiara,
  Paternostro, and Palma}}]{de2011entanglement}
\bibinfo{author}{\bibfnamefont{G.}~\bibnamefont{De~Chiara}},
  \bibinfo{author}{\bibfnamefont{M.}~\bibnamefont{Paternostro}},
  \bibnamefont{and} \bibinfo{author}{\bibfnamefont{G.~M.} \bibnamefont{Palma}},
  \bibinfo{journal}{, Phys. Rev. A} \textbf{\bibinfo{volume}{83}},
  \bibinfo{pages}{052324} (\bibinfo{year}{2011}).

\bibitem[{\citenamefont{Yin et~al.}(2013)\citenamefont{Yin, Li, Zhang, and
  Duan}}]{yin2013large}
\bibinfo{author}{\bibfnamefont{Z.-q.} \bibnamefont{Yin}},
  \bibinfo{author}{\bibfnamefont{T.}~\bibnamefont{Li}},
  \bibinfo{author}{\bibfnamefont{X.}~\bibnamefont{Zhang}}, \bibnamefont{and}
  \bibinfo{author}{\bibfnamefont{L.}~\bibnamefont{Duan}}, \bibinfo{journal}{,
  Phys. Rev. A} \textbf{\bibinfo{volume}{88}}, \bibinfo{pages}{033614}
  (\bibinfo{year}{2013}).

\bibitem[{\citenamefont{Nunnenkamp et~al.}(2011)\citenamefont{Nunnenkamp,
  B{\o}rkje, and Girvin}}]{nunnenkamp2011single}
\bibinfo{author}{\bibfnamefont{A.}~\bibnamefont{Nunnenkamp}},
  \bibinfo{author}{\bibfnamefont{K.}~\bibnamefont{B{\o}rkje}},
  \bibnamefont{and} \bibinfo{author}{\bibfnamefont{S.}~\bibnamefont{Girvin}},
  \bibinfo{journal}{, Phys. Rev. Lett.} \textbf{\bibinfo{volume}{107}},
  \bibinfo{pages}{063602} (\bibinfo{year}{2011}).

\bibitem[{\citenamefont{Liao et~al.}(2012)\citenamefont{Liao, Cheung, and
  Law}}]{liao2012spectrum}
\bibinfo{author}{\bibfnamefont{J.-Q.} \bibnamefont{Liao}},
  \bibinfo{author}{\bibfnamefont{H.}~\bibnamefont{Cheung}}, \bibnamefont{and}
  \bibinfo{author}{\bibfnamefont{C.}~\bibnamefont{Law}}, \bibinfo{journal}{,
  Phys. Rev. A} \textbf{\bibinfo{volume}{85}}, \bibinfo{pages}{025803}
  (\bibinfo{year}{2012}).

\bibitem[{\citenamefont{Mirza and van Enk}(2014)}]{imran2014single}
\bibinfo{author}{\bibfnamefont{I.~M.} \bibnamefont{Mirza}} \bibnamefont{and}
  \bibinfo{author}{\bibfnamefont{S.}~\bibnamefont{van Enk}},
  \bibinfo{journal}{, Phys. Rev. A} \textbf{\bibinfo{volume}{90}},
  \bibinfo{pages}{043831} (\bibinfo{year}{2014}).

\bibitem[{\citenamefont{Mirza}(2015)}]{mirza2015real}
\bibinfo{author}{\bibfnamefont{I.~M.} \bibnamefont{Mirza}}, \bibinfo{journal}{,
  JOSA B} \textbf{\bibinfo{volume}{32}}, \bibinfo{pages}{1604}
  (\bibinfo{year}{2015}).

\bibitem[{\citenamefont{Walls}(1970)}]{walls1970higher}
\bibinfo{author}{\bibfnamefont{D.~F.} \bibnamefont{Walls}}, \bibinfo{journal}{,
  Zeitschrift f{\"u}r Physik} \textbf{\bibinfo{volume}{234}},
  \bibinfo{pages}{231} (\bibinfo{year}{1970}).

\bibitem[{\citenamefont{Cresser}(1992)}]{cresser1992thermal}
\bibinfo{author}{\bibfnamefont{J.}~\bibnamefont{Cresser}}, \bibinfo{journal}{,
  J. Mod. Opt.} \textbf{\bibinfo{volume}{39}}, \bibinfo{pages}{2187}
  (\bibinfo{year}{1992}).

\bibitem[{\citenamefont{Joshi et~al.}(2014)\citenamefont{Joshi, {\"O}hberg,
  Cresser, and Andersson}}]{joshi2014markovian}
\bibinfo{author}{\bibfnamefont{C.}~\bibnamefont{Joshi}},
  \bibinfo{author}{\bibfnamefont{P.}~\bibnamefont{{\"O}hberg}},
  \bibinfo{author}{\bibfnamefont{J.~D.} \bibnamefont{Cresser}},
  \bibnamefont{and}
  \bibinfo{author}{\bibfnamefont{E.}~\bibnamefont{Andersson}},
  \bibinfo{journal}{, Phys. Rev. A} \textbf{\bibinfo{volume}{90}},
  \bibinfo{pages}{063815} (\bibinfo{year}{2014}).

\bibitem[{\citenamefont{Nicolosi et~al.}(2004)\citenamefont{Nicolosi, Napoli,
  Messina, and Petruccione}}]{nicolosi2004dissipation}
\bibinfo{author}{\bibfnamefont{S.}~\bibnamefont{Nicolosi}},
  \bibinfo{author}{\bibfnamefont{A.}~\bibnamefont{Napoli}},
  \bibinfo{author}{\bibfnamefont{A.}~\bibnamefont{Messina}}, \bibnamefont{and}
  \bibinfo{author}{\bibfnamefont{F.}~\bibnamefont{Petruccione}},
  \bibinfo{journal}{, Phys. Rev. A} \textbf{\bibinfo{volume}{70}},
  \bibinfo{pages}{022511} (\bibinfo{year}{2004}).

\bibitem[{\citenamefont{Birnbaum et~al.}(2005)\citenamefont{Birnbaum, Boca,
  Miller, Boozer, Northup, and Kimble}}]{birnbaum2005photon}
\bibinfo{author}{\bibfnamefont{K.~M.} \bibnamefont{Birnbaum}},
  \bibinfo{author}{\bibfnamefont{A.}~\bibnamefont{Boca}},
  \bibinfo{author}{\bibfnamefont{R.}~\bibnamefont{Miller}},
  \bibinfo{author}{\bibfnamefont{A.~D.} \bibnamefont{Boozer}},
  \bibinfo{author}{\bibfnamefont{T.~E.} \bibnamefont{Northup}},
  \bibnamefont{and} \bibinfo{author}{\bibfnamefont{H.~J.}
  \bibnamefont{Kimble}}, \bibinfo{journal}{, Nature}
  \textbf{\bibinfo{volume}{436}}, \bibinfo{pages}{87} (\bibinfo{year}{2005}).

\bibitem[{\citenamefont{Guerlin et~al.}(2007)\citenamefont{Guerlin, Bernu,
  Deleglise, Sayrin, Gleyzes, Kuhr, Brune, Raimond, and
  Haroche}}]{guerlin2007progressive}
\bibinfo{author}{\bibfnamefont{C.}~\bibnamefont{Guerlin}},
  \bibinfo{author}{\bibfnamefont{J.}~\bibnamefont{Bernu}},
  \bibinfo{author}{\bibfnamefont{S.}~\bibnamefont{Deleglise}},
  \bibinfo{author}{\bibfnamefont{C.}~\bibnamefont{Sayrin}},
  \bibinfo{author}{\bibfnamefont{S.}~\bibnamefont{Gleyzes}},
  \bibinfo{author}{\bibfnamefont{S.}~\bibnamefont{Kuhr}},
  \bibinfo{author}{\bibfnamefont{M.}~\bibnamefont{Brune}},
  \bibinfo{author}{\bibfnamefont{J.-M.} \bibnamefont{Raimond}},
  \bibnamefont{and} \bibinfo{author}{\bibfnamefont{S.}~\bibnamefont{Haroche}},
  \bibinfo{journal}{, Nature} \textbf{\bibinfo{volume}{448}},
  \bibinfo{pages}{889} (\bibinfo{year}{2007}).

\bibitem[{\citenamefont{Alharbi and Ficek}(2010)}]{alharbi2010deterministic}
\bibinfo{author}{\bibfnamefont{A.}~\bibnamefont{Alharbi}} \bibnamefont{and}
  \bibinfo{author}{\bibfnamefont{Z.}~\bibnamefont{Ficek}}, \bibinfo{journal}{,
  Phys. Rev. A} \textbf{\bibinfo{volume}{82}}, \bibinfo{pages}{054103}
  (\bibinfo{year}{2010}).

\bibitem[{\citenamefont{Hu et~al.}(2015)\citenamefont{Hu, Huang, Liao, Tian,
  and Goan}}]{hu2015quantum}
\bibinfo{author}{\bibfnamefont{D.}~\bibnamefont{Hu}},
  \bibinfo{author}{\bibfnamefont{S.-Y.} \bibnamefont{Huang}},
  \bibinfo{author}{\bibfnamefont{J.-Q.} \bibnamefont{Liao}},
  \bibinfo{author}{\bibfnamefont{L.}~\bibnamefont{Tian}}, \bibnamefont{and}
  \bibinfo{author}{\bibfnamefont{H.-S.} \bibnamefont{Goan}},
  \bibinfo{journal}{, Phys. Rev. A} \textbf{\bibinfo{volume}{91}},
  \bibinfo{pages}{013812} (\bibinfo{year}{2015}).

\bibitem[{\citenamefont{Holz et~al.}(2015)\citenamefont{Holz, Betzholz, and
  Bienert}}]{holz2015suppression}
\bibinfo{author}{\bibfnamefont{T.}~\bibnamefont{Holz}},
  \bibinfo{author}{\bibfnamefont{R.}~\bibnamefont{Betzholz}}, \bibnamefont{and}
  \bibinfo{author}{\bibfnamefont{M.}~\bibnamefont{Bienert}},
  \bibinfo{journal}{. Phys. Rev. A} \textbf{\bibinfo{volume}{92}},
  \bibinfo{pages}{043822} (\bibinfo{year}{2015}).

\bibitem[{\citenamefont{Eberly and Wodkiewicz}(1977)}]{eberly1977time}
\bibinfo{author}{\bibfnamefont{J.}~\bibnamefont{Eberly}} \bibnamefont{and}
  \bibinfo{author}{\bibfnamefont{K.}~\bibnamefont{Wodkiewicz}},
  \bibinfo{journal}{, JOSA} \textbf{\bibinfo{volume}{67}},
  \bibinfo{pages}{1252} (\bibinfo{year}{1977}).

\bibitem[{\citenamefont{Zhang et~al.}(2014)\citenamefont{Zhang, Tan, and
  Barker}}]{zhang2014effects}
\bibinfo{author}{\bibfnamefont{Y.-Q.} \bibnamefont{Zhang}},
  \bibinfo{author}{\bibfnamefont{L.}~\bibnamefont{Tan}}, \bibnamefont{and}
  \bibinfo{author}{\bibfnamefont{P.}~\bibnamefont{Barker}}, \bibinfo{journal}{,
  Phys. Rev. A} \textbf{\bibinfo{volume}{89}}, \bibinfo{pages}{043838}
  (\bibinfo{year}{2014}).

\end{thebibliography}

\end{document}